\documentclass{gistt}

\usepackage{makecell}
\usepackage{graphicx}
\usepackage{float}
\usepackage{svg}
\usepackage{siunitx}
\usepackage{dblfloatfix}
\usepackage{subcaption}
\usepackage{tablefootnote}
\usepackage{scrextend}

\DeclareSIUnit{\minute}{m}

\makeatletter
\newcommand{\sitime}[1]{\sitime@aux#1:::\@nil}
\def\sitime@aux#1:#2:#3:#4\@nil{%
  \@ifmtarg{#1#2#3}%
  {%
    \PackageError{siunitx (modified)}%
                 {Empty \string\sitime\space argument}{}%
  }%
{%
\@ifmtarg{#1}{}{\SI{#1}{\hour}\@ifmtarg{#2#3}{}{\csname l_siunitx_number_unit_separator_tl\endcsname}}%
\@ifmtarg{#2}{}{\SI{#2}{\minute}\@ifmtarg{#3}{}{\csname l_siunitx_number_unit_separator_tl\endcsname}}%
\@ifmtarg{#3}{}{\SI{#3}{\second}}%
}%
}
\makeatother

\title{Dynamic and Static Analysis of Python Software with Kieker Including Reconstructed Architectures}
\author{Daphné Larrivain\\
daphne.larrivain@ecole.ensicaen.fr\\
ENSICAEN, Caen, France
\and
Shinhyung Yang\\
shinhyung.yang@email.uni-kiel.de\\
Kiel University, Kiel, Germany
\and
Wilhelm Hasselbring\\
hasselbring@email.uni-kiel.de\\
Kiel University, Kiel, Germany}

\begin{document}

\maketitle

\begin{abstract}

The Kieker observability framework is a tool that provides users with the means to design a custom observability pipeline for their application. Originally tailored for Java, supporting Python with Kieker is worthwhile. Python's popularity has exploded over the years, thus making structural insights of Python applications highly valuable. Our Python analysis pipeline combines static and dynamic analysis in order to build a complete picture of a given system.

\end{abstract}

\section{Introduction}

Reverse engineering is the cornerstone of maintainability. Visual access to a system’s structure enables valuable insights, from identifying differences between intended and implemented architectures to understanding unfamiliar software.

Static analysis extracts key insights from source code, whereas dynamic analysis uncovers runtime behaviour. Combining the two allows for a more comprehensive view. The latter is known as combined 
analysis. This approach builds on earlier research.

Using the Kieker observability framework~\cite{Hasselbring:2020, Yang:2025} as a basis, the existing tools were extended to bring a custom analysis and visualization pipeline to Python.

\section{Related Work}

There are many types of static analysis tools. Amongst them are AST matchers~\cite{Gulabovska:2019}. An abstract syntax tree~(AST) is a tree-like structure that represents the syntactic organization of source code, where each node corresponds to the associated code construct. According to the characteristics of simple and flat Python development, ASTs are used as the main means to mine more context information of Python code fragments~\cite{Ma:2022}.




In terms of architecture recovery, Arcade~\cite{Laser:2020} provides limited Python support. It aims to prevent architectural decay through a modular pipeline with five subsystems: recovery, decay detection, measurement, visualization, and prediction. Arcade doesn’t mention the use of dynamic analysis. Unlike Arcade, we focus on the combined analysis of Python software, especially on visualization.

Kieker’s existing approach, originally designed for Fortran~\cite{SSP2022Fortran}, offers principles adaptable to Python. The Fortran workflow involved three steps: Fxtran translates target source code into an AST,\tablefootnote{\url{https://github.com/pmarguinaud/fxtran}} Fxca converted the AST to a CSV format compatible with SAR, and SAR generated an architectural model from the CSV.

In this paper, we present two main contributions. First, we adapted the existing combined analysis approach, originally developed for Fortran applications, to effectively analyze Python applications. Second, we incorporated the Tulip framework into our workflow, which significantly enhances visualization performance.

\section{Implementation}

\subsection{General Pipeline}


\begin{figure*}[htbp]
\centering
\includegraphics[width=\linewidth]{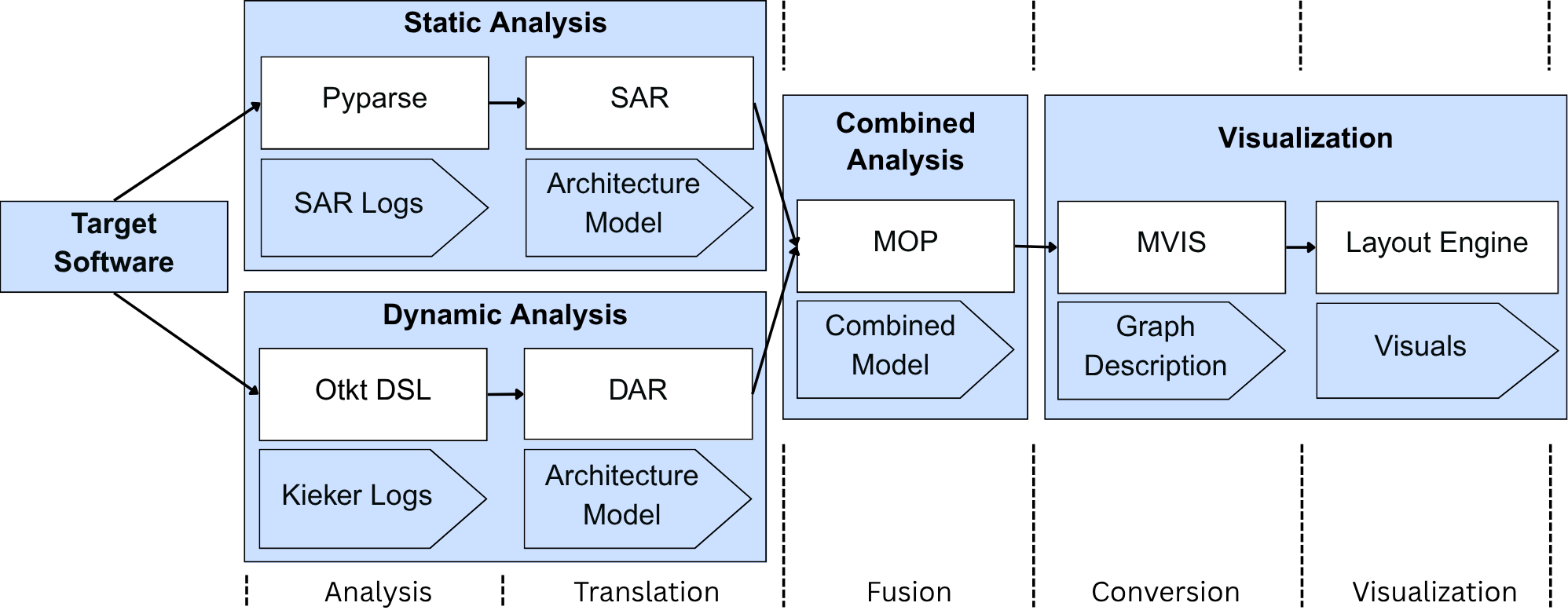}
\caption{Combined analysis pipeline stages. Filter tools are shown in white, the intermediate data formats are shown in green, and the different stages are in white/gray.}
\label{fig:CombinedAnalysisPipeline}
\end{figure*}


\begin{table*}[t]
\centering
\begin{tabular}{r p{7.753cm} r r p{1.823cm}} 
 \textbf{Tool} & \textbf{Description, repository} & \textbf{Origin} & \textbf{Revised} & \textbf{Language} \\
\hline
Pyparse\tablefootnote{\url{https://github.com/kieker-monitoring/pyparse}} & analyzes Python code and outputs AST for SAR & New & New & Python \\
\hline
OtktInst & (Otkt Instrument tool)\tablefootnote{\url{https://github.com/kieker-monitoring/OtktInst}} applies Otkt DSL to instrument Python code & New & New & Python \\
\hline
GGVIS & (Grouped Graph Visualizer)\tablefootnote{\url{https://github.com/kieker-monitoring/GGVIS}} renders graphs and exports to PDF/SVG/PNG & New & New & Python \\
\hline
Otkt DSL & (OpenTelemetry to Kieker Translation DSL)\tablefootnote{\url{https://github.com/kieker-monitoring/OtktDSL}} describes a mapping from OTel span to Kieker record & Kieker & Improved & Python, Java, Xtext \\
\hline
DAR & (Dynamic Architecture Recovery)\tablefootnote{\url{https://github.com/kieker-monitoring/kieker}\label{KiekerRepo}} converts Kieker logs to architecture models & Kieker & As-is & Java \\
\hline
SAR & (Static Architecture Recovery)\footref{KiekerRepo} builds architecture models from received AST description & Kieker & Fixes & Java \\
\hline
MOP & (Model OPeration)\footref{KiekerRepo} merges and compares architecture models & Kieker & As-is & Java \\
\hline
MVIS & (Model Visualization and Statistics Tool)\footref{KiekerRepo} exports architecture models to grphics & Kieker & As-is & Java \\
\end{tabular}
\caption{We developed three new tools and revised two Kieker tools for our combined analysis pipeline.}
\label{tab:tools}
\end{table*}

The Kieker observability framework offers a broad set of tools. They follow the TeeTime pipes and filters architectural pattern~\cite{Wulf:2017}. In this model, each filter represents a processing step that transforms incoming data into output data. Pipes serve as connections between these steps, enabling data to flow from one filter tool to the next. For tools to be chained together, the output format of one must match the input format of the next. When this condition is met, users can combine filters to build custom processing pipelines and generate the desired output.

Since the Kieker framework was not originally designed to handle Python
applications, it was necessary to adapt existing tools and create new ones as
described in Table~\ref{tab:tools}. The resulting processing pipeline is shown in Figure~\ref{fig:CombinedAnalysisPipeline}.

\subsection{Dynamic Analysis}

Our proposed pipeline begins with dynamic analysis. Kieker's Python support~\cite{Simonov:2023} is based on OpenTelemetry, a language- and platform-agnostic observability framework. Thus, Kieker uses OtktDSL~\cite{Simonov:2024} to translate OpenTelemetry spans into Kieker records, which allow for Kieker's dynamic analysis.

This bridge is technical, not semantic. It enables data transfer but leaves interpretation and data selection to the user. To ensure tool compatibility, a Java-inspired architecture was adopted. The reconstruction considers not only files but also classes, adding an object-oriented layer to the system’s representation.

Analyzing a specific application requires instrumentation tailored to the desired insights. While this varies per case, semi-automation is possible using naming-convention heuristics. Otkt-Instrument has been developed for this purpose. The obtained Kieker logs can then be given to DAR to obtain the associated architecture model.

\subsection{Static Analysis}

Subsequently, we turn to static analysis. Kieker’s existing approach,
originally designed for Fortran~\cite{SSP2022Fortran}, offers principles
adaptable to Python. The Fortran workflow involved three steps:
Fxtran\tablefootnote{\url{https://github.com/pmarguinaud/fxtran}} translates
code into AST, fxca converts AST insights to the CSV format compatible with SAR, and SAR generates
architectural model from the received AST.

This outlines a general process: source code → AST → CSV → SAR tool. However, this sequence couldn’t be reused directly for Python. ASTs are language-specific, and Python’s built-in module doesn’t produce output compatible with existing Kieker tools. To solve this, Pyparse (see Table~\ref{tab:tools}) was built. It performs AST generation and CSV conversion in one step, extracting two key types of relationships: function calls and data flow. These are then passed to the SAR tool for architecture reconstruction.

\subsection{Combined Analysis}


With both dynamic and static analyses in place, the next challenge was merging their outputs into a single architectural model. A deeper inspection of the DAR and SAR tools revealed inconsistencies in how each one interpreted and used the architectural model. Even minor differences in semantics could cause the final model to remain fragmented, with the DAR and SAR parts isolated from each other. To bridge these differences in naming conventions, a technical fix of the SAR tool was developed to match the DAR tool. Ideally, the SAR fix would have been implemented directly within the tool. However, due to the lack of working source code, standardizing its behavior was not feasible.

The SustainKieker project emerged in response to challenges around long-term maintainability. The Kieker tool suite reflects the typical lifecycle of research software: it evolves alongside academic projects, often shaped by the needs of individual theses. However, this also means that contributions are sometimes treated more as functional prototypes than fully integrated components.

As a result, several tools are currently non-functional in the official repository, and documentation remains limited. Much of the technical insight is embedded in replication packages or scattered across related publications. Despite these hurdles, the tools could still be used by tracing older, functional versions from previous replication packages. Ultimately, an archive containing working binaries was located.


\subsection{Visualization}


\begin{figure*}[htbp]
\centering
\includegraphics[width=\linewidth]{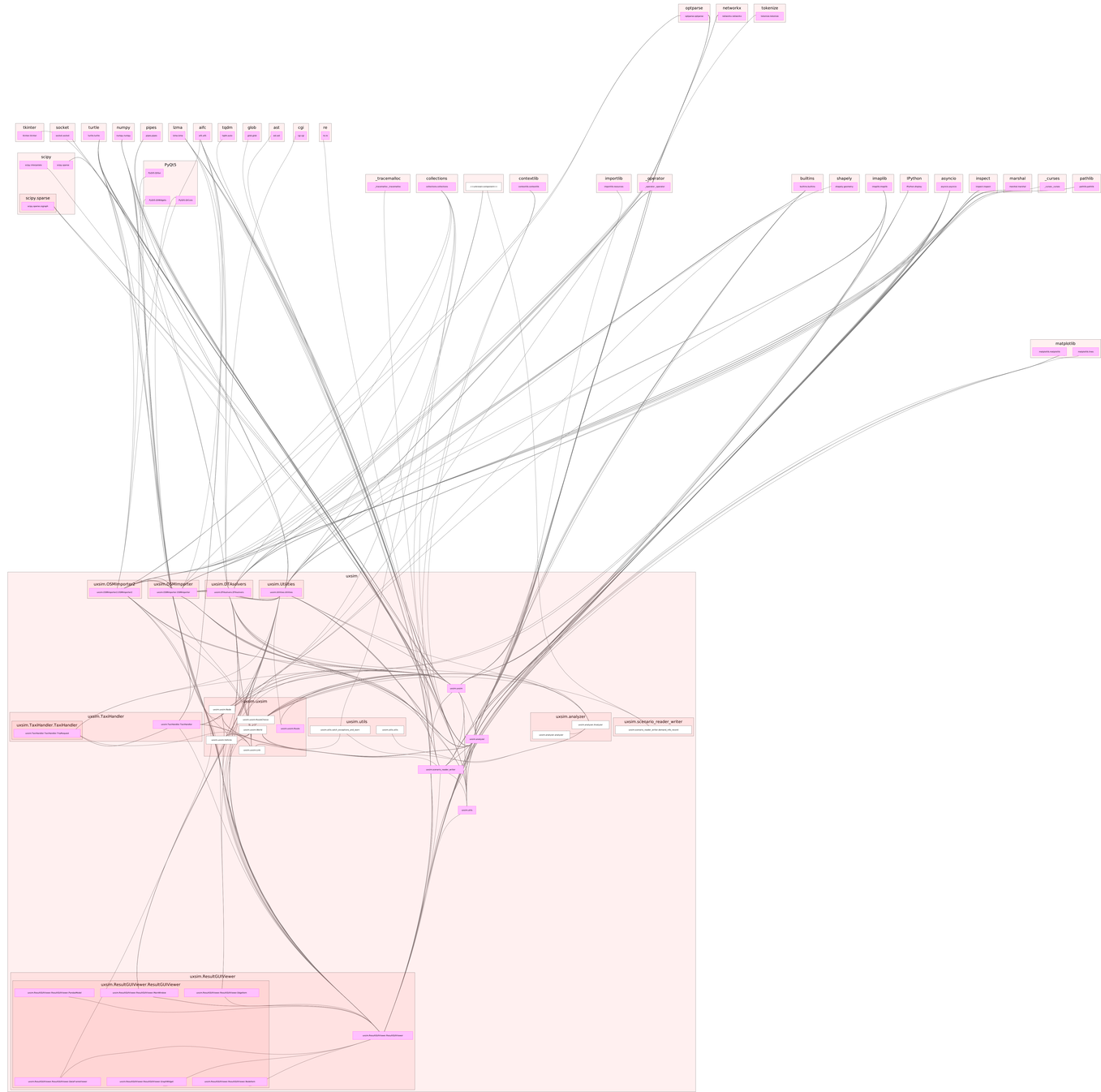}
\caption{UXSim, one of the evaluated applications, grouped with Tulip. The \emph{++ unknown component ++} is a relic from the dynamic analysis entry point.}
\label{fig:hello}
\end{figure*}


At last comes the visualization. MVIS (see Table~\ref{tab:tools}) was first used. It converts specific elements of the architectural model into standardized graph description formats. This step defines the structure of the graph but does not handle its layout.


A key objective of the visualization was to lay out the graph to reveal the system’s structure at a glance. The displayed components were flattened in the default output, without any meaningful grouping, which made the result difficult to interpret. To improve readability, nodes needed to be grouped according to their package affiliation.

Graphviz, the default visualization toolkit, presents notable limitations in handling nested graph layouts~\cite{Samodra:2025}. Among its suite of layout engines, only dot and fdp support nested structures. The dot engine employs a hierarchical layout that yields cleaner diagrams but struggles to scale with increasing graph complexity. Conversely, fdp uses a force-directed approach that accommodates larger graphs, though it often suffers from excessive edge crossings. Unfortunately, both engines reach scalability limits when faced with highly complex architectures.

To overcome these constraints, we developed GGVIS\ref{tab:tools}, a custom tool built on the Tulip framework~\cite{Auber:2017}. The alternative to Graphviz needed to support nested structures, either natively or through flexible configuration, while also allowing scripting and compatibility with standard formats such as GraphML or, preferably, DOT.

As previously discussed, the MVIS tool translates architecture models into graph description formats, generating two distinct outputs: DOT and GraphML. These formats differ in granularity. The DOT output separates components and operations, whereas the GraphML output embeds operations within their respective components. In other words, the two formats are not interchangeable. For producing a clean component diagram with minimal post-processing, DOT was the preferred format.

One of the main challenges in selecting a new visualization tool was that most existing solutions are either Graphviz-based, GUI-centric, or web-based. Tulip, however, met all the necessary criteria: it supports nesting, offers scripting capabilities, and handles both DOT and GraphML formats effectively.

\begin{figure*}[htbp]
  \centering

  \begin{subfigure}[t]{0.48\textwidth}
    \centering
    \includegraphics[width=\linewidth]{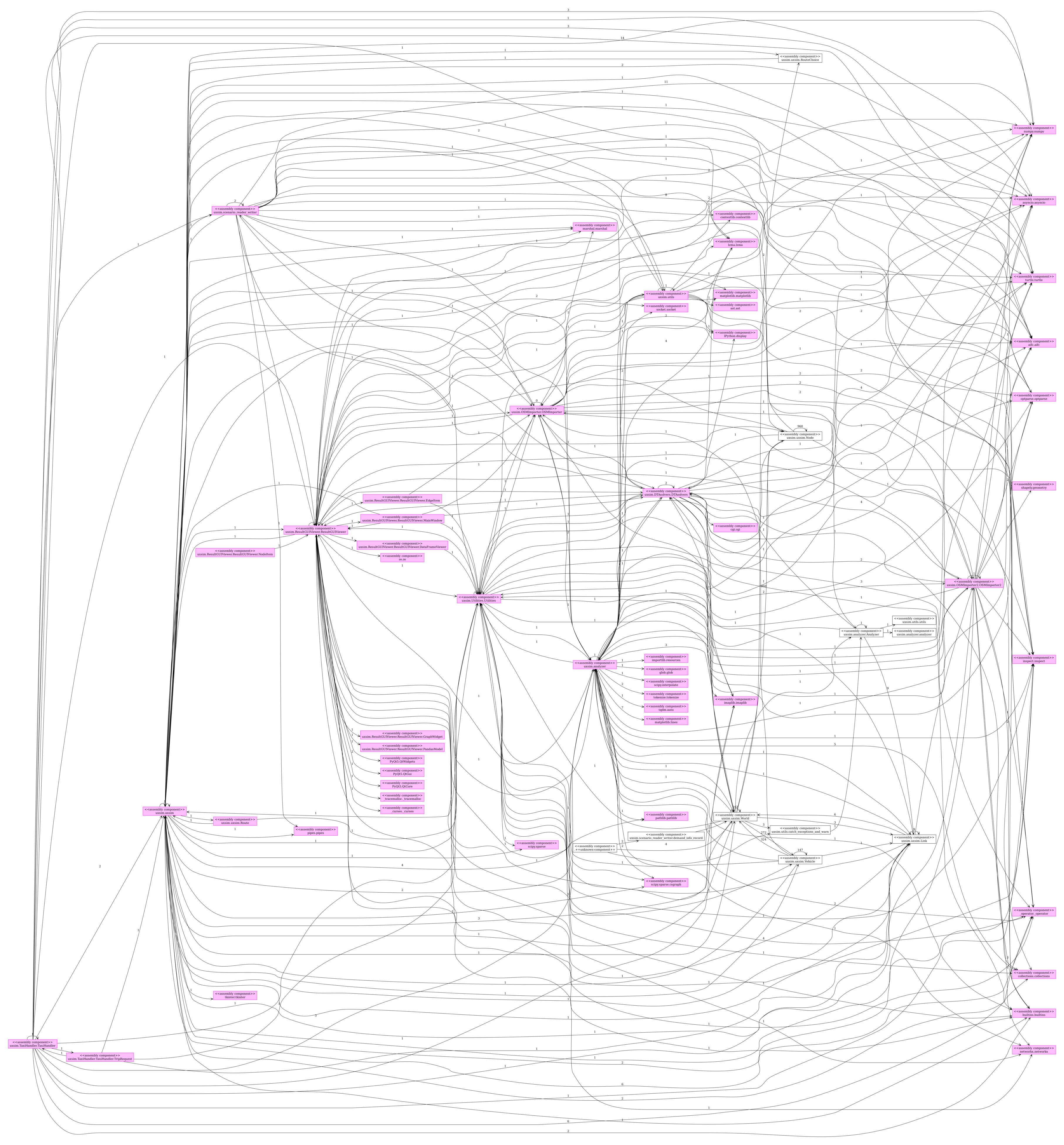}
    \caption{Flat graph with dot}
    \label{fig:uxsim-a}
  \end{subfigure}
  \hfill
  \begin{subfigure}[t]{0.48\textwidth}
    \centering
    \includegraphics[width=\linewidth]{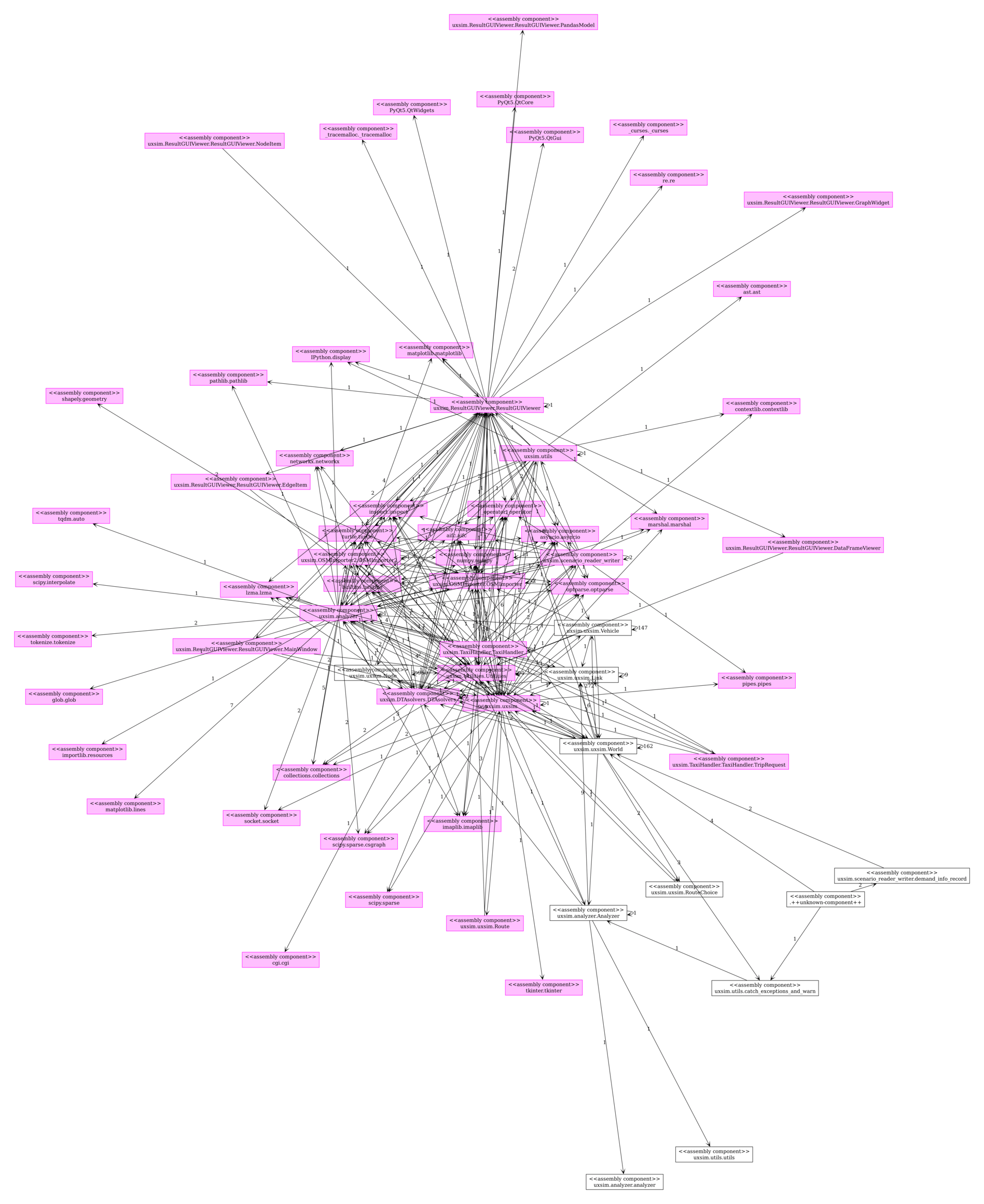}
    \caption{Flat graph with fdp}
    \label{fig:uxsim-b}
  \end{subfigure}

  \vspace{1em}

  \begin{subfigure}[t]{\textwidth}
    \centering
    \includegraphics[width=\linewidth]{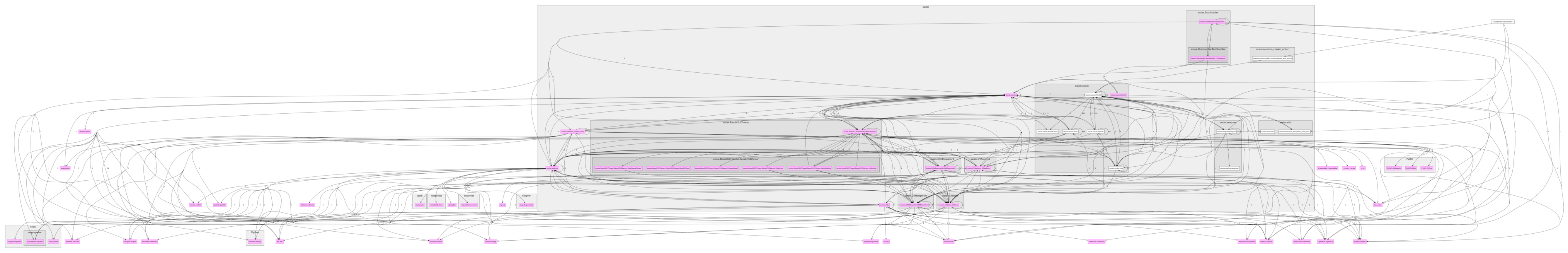}
    \caption{Grouped graph with dot}
    \label{fig:uxsim-c}
  \end{subfigure}

  \vspace{1em}

  \begin{subfigure}[t]{0.48\textwidth}
    \centering
    \includegraphics[width=\linewidth]{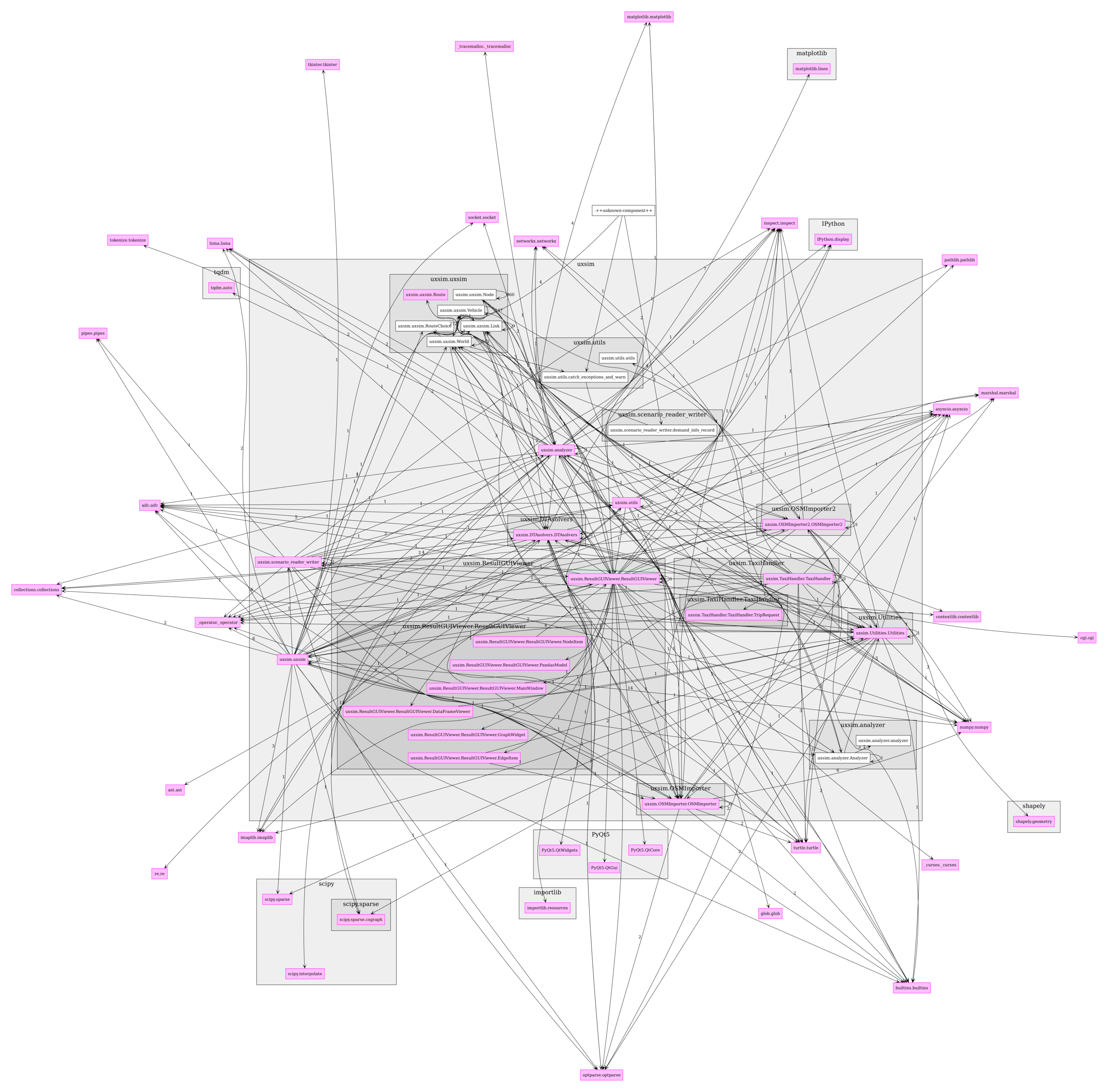}
    \caption{Grouped graph with fdp}
    \label{fig:uxsim-d}
  \end{subfigure}
  \hfill
  \begin{subfigure}[t]{0.48\textwidth}
    \centering
    \includegraphics[width=\linewidth]{img/uxsim_grouped_ggvis.png}
    \caption{Grouped graph with GGVIS}
    \label{fig:uxsim-e}
  \end{subfigure}

  \caption{Visualization of different layout variants for the combined UXsim graph. The version without test components was selected to ensure compatibility with Graphviz’s dot engine, as the full graph exceeds its processing capabilities. All layouts were generated from the same underlying graph description.}
  \label{fig:uxsim-process}
\end{figure*}

\begin{figure*}[htbp]
\centering
\includegraphics[width=\linewidth]{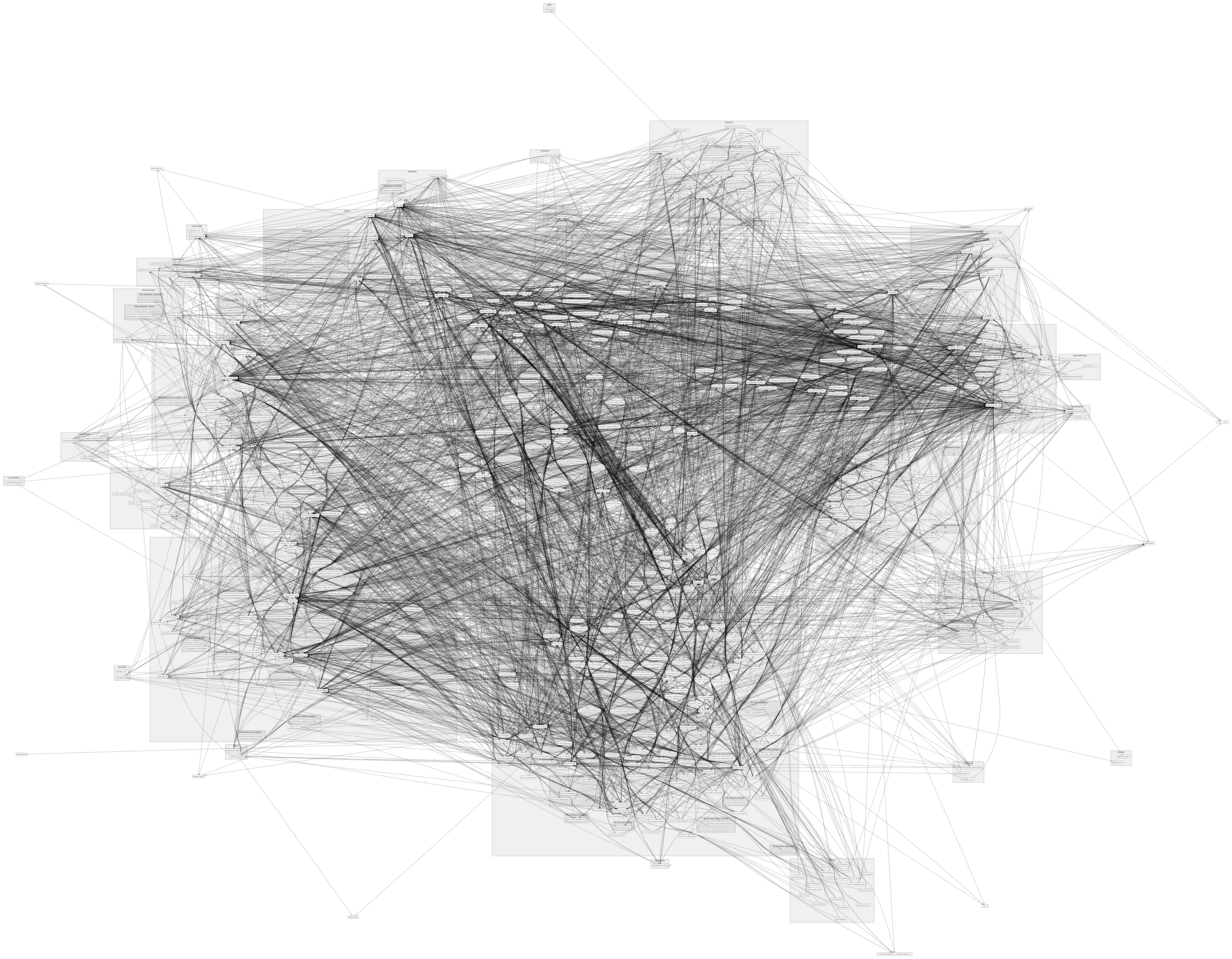}
\caption{Numpy with fdp.}
\label{fig:numpy-fdp}
\end{figure*}

\begin{figure*}[htbp]
\centering
\includegraphics[width=0.68\linewidth]{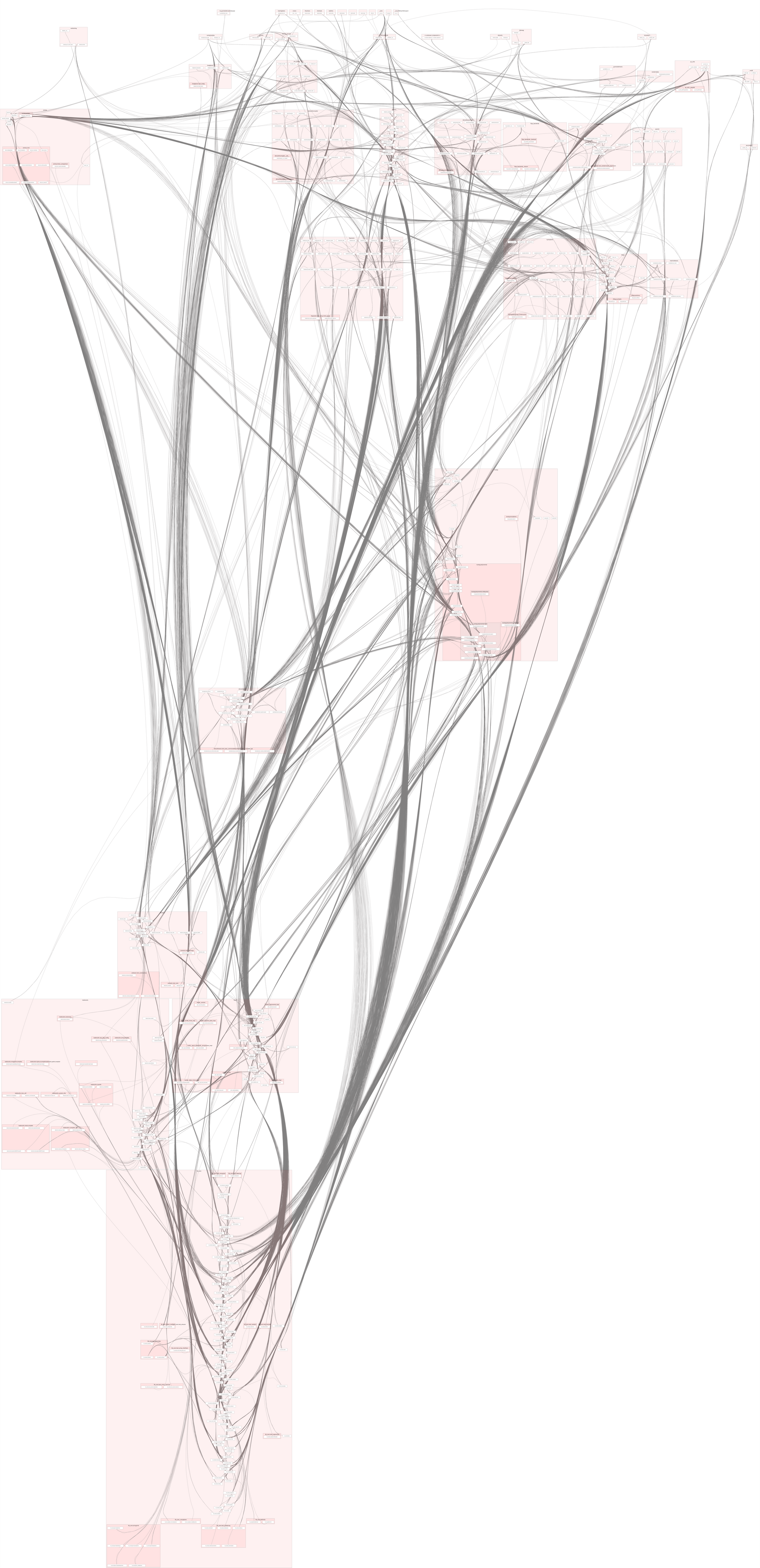}
\caption{Numpy with ggvis.}
\label{fig:numpy-ggvis}
\end{figure*}

For example, figures \ref{fig:numpy-fdp} and \ref{fig:numpy-ggvis} illustrate the NumPy architecture rendered using the fdp and GGVIS layout engines, respectively. The rendering performance has improved significantly: fdp required \sitime{:16:38}, whereas GGVIS completes the task in approximately \sitime{::30} to \sitime{:1:00}, depending on the desired output format (e.g., PDF, SVG, PNG).

\subsection{Evaluation}

\subsubsection{General Pipeline Evaluation}

The pipeline was tested on several Python projects of various sizes.

We measured the duration of the entire pipeline, except for the data gathering phase of the dynamic analysis. This step was excluded because it includes the runtime overhead of the application under test, not the performance of the pipeline itself.


\begin{table*}[t]
\centering
\resizebox{\textwidth}{!}{
\begin{tabular}{||l|@{\hskip 4pt}c|@{\hskip 4pt}c|@{\hskip 4pt}c|@{\hskip 4pt}c|@{\hskip 4pt}c|@{\hskip 4pt}c||} 
 \hline
 \textbf{} & \textbf{Anytree} & \textbf{UXsim} & \textbf{Pillow} & \textbf{Matplotlib} & \textbf{Numpy} & \textbf{Scipy} \\
 \hline\hline
Pyparse & \sitime{::1}  & \sitime{::2} & \sitime{::10} & \sitime{:3:53} & \sitime{:2:57} & \sitime{:7:15} \\
DAR     & \sitime{::2}  & \sitime{::4} & \sitime{::1}  & \sitime{::2}   &                 & \sitime{:0:02} \\
SAR     & \sitime{::3}  & \sitime{::6} & \sitime{::10} & \sitime{:7:26} & \sitime{:10:44} & \sitime{:21:58} \\
MOP     & \sitime{::3}  & \sitime{::6} & \sitime{::5}  & \sitime{:26:13}&                 & \sitime{::28} \\
MVIS    & \sitime{::2}  & \sitime{::3} & \sitime{::2}  & \sitime{:1:10} & \sitime{:19:22} & \sitime{::3} \\
GGVIS   & \sitime{::1}  & \sitime{::1} & \sitime{::1}  & \sitime{:4:28} & \sitime{::34}   & \sitime{::1} \\
 \hline\hline
Total   & \sitime{::12} & \sitime{::22} & \sitime{::29} & \sitime{:43:12} & \sitime{:33:37} & \sitime{:29:47} \\
 \hline
\end{tabular}
}
\caption{Pipeline durations in minutes and seconds.}
\label{tab:pipeline-results}
\end{table*}


In Table~\ref{tab:pipeline-results}, a few observations deserve to be highlighted. First, as the number of files increases, processing time increases accordingly.

Then, instrumentation succeeded for Anytree\tablefootnote{\url{https://github.com/c0fec0de/anytree}}, UXsim~\cite{Seo:2025}, and Pillow\tablefootnote{\url{https://github.com/python-pillow/Pillow}} without issues.

Matplotlib\tablefootnote{\url{https://github.com/matplotlib/matplotlib}} and Scipy\tablefootnote{\url{https://github.com/scipy/scipy}} show low DAR times due to reduced data input. By design, dynamic analysis is more focused than static analysis. However, Numpy couldn't be instrumented due to application-level limitations. Thus, the dynamic analysis wasn't possible.

Finally, regarding Scipy, the MOP tool failed to process the static model. Despite this, DAR and SAR models were structurally valid and produced correct graphs via MVIS and GVIS. This explains the low processing time.


\subsubsection{Visualization Evaluation}

\begin{table}[H]
\centering
\resizebox{\textwidth/2 - 10pt}{!}{
\begin{tabular}{||l|@{\hskip 4pt}c|@{\hskip 4pt}c|@{\hskip 4pt}c||} 
 \hline
 \textbf{} & \textbf{UXsim} & \textbf{Anytree} & \textbf{Numpy} \\
 \hline\hline
Fdp          & \sitime{::3}      & \sitime{::2}     & \sitime{:1:50}   \\
Dot          & \sitime{:1:09}**  & \sitime{::18}    & $\infty$         \\
Fdp grouped  & \sitime{::6}      & \sitime{::2}     & \sitime{:16:36}  \\
Dot grouped  & $\infty$          & $\infty$         & $\infty$         \\
GGVIS        & \sitime{::1}      & \sitime{::1}     & \sitime{::34}    \\
 \hline
\end{tabular}
}
\caption{Execution times for different layout engines.}
\label{tab:visu-times}
\end{table}

Graphviz’s Dot and FDP layouts were compared for visualizing flat and nested graphs, Dot and FDP being the only Graphviz tools supporting nesting, as well as the new solution GGVIS\ref{tab:visu-times}. As the graph size increases, Dot shows performance issues. Fdp holds up better, but still shows great computing times when it comes to Numpy.


\section{Conclusion}

This work extends the Kieker observability framework to improve its support for
Python applications. A combined analysis approach, integrating static and
dynamic analysis, was
replicated.\footnote{\url{https://zenodo.org/records/16735614}} To implement this, a custom analysis pipeline was designed using built-in Kieker tools, alongside adaptations of existing components and the development of new ones.

Throughout the process, inconsistencies in tool outputs and maintenance challenges were identified. To resolve these issues, solutions from prior research were customized and integrated to better align with the project’s requirements. A Python static analyzer based on AST matching was developed specifically for the pipeline. Furthermore, a more intuitive visualization method was introduced, and limitations in current layout tools were revealed, highlighting an area for future improvement.

\paragraph{Acknowledgment}
This research is funded by the Deutsche Forschungsgemeinschaft (DFG -- German
Research Foundation), grant no.~528713834.

\printbibliography

\end{document}